\documentclass{tMPH2e}
\usepackage{longtable}
\usepackage{graphicx}     
\usepackage{lscape}
\usepackage{color}
\usepackage{amsmath}

\newcommand{\etal}{{\it et al.}}
\newcommand{\ai}{{\it ab initio }}

\newcommand{\cm}{cm$^{-1}$}

\newcommand{\hp}{H$_3^+$}

\def\a0{{$a_{\rm 0}$}}

\begin{document}
\articletype{ARTICLE}

\title{A Global  Potential Energy Surface for H$_3^+$}

\author{
Irina I. Mizus$^{1}$,   Oleg L. Polyansky,$^{1,2,\dag}$\footnote{$\dag$ o.polyansky@ucl.ac.uk}
Laura K. McKemmish$^{2,3}$
Jonathan Tennyson,$^{2,*}$,\footnote{* j.tennyson@ucl.ac.uk} 
Alexander Alijah$^{4}$ and
Nikolai F. Zobov$^{1}$,\\  
$^{1}$Institute of Applied Physics, Russian Academy of Sciences,
Ulyanov Street 46, Nizhny Novgorod, Russia 603950\\
$^{2}$Department of Physics and Astronomy, University College London, Gower Street,  London WC1E~6BT, UK\\
$^{3}$Department of Chemistry, University of New South Wales, Australia\\
$^{4}$Groupe de Spectrom\'etrie Mol\'eculaire et Atmosph\'erique,
GSMA, UMR CNRS 7331, Universit\'e de Reims Champagne-Ardenne, France}

\date{\today}

\maketitle

\begin{abstract}
A globally correct potential energy surface (PES) for the \hp\ 
molecular ion is presented. The  Born-Oppenheimer (BO) \ai\
grid points of Pavanello \etal\ 
[\textit{J. Chem. Phys.} {\bf  136}, 184303 (2012)]
are refitted as BOPES75K, which  reproduces the energies below dissociation 
with a root mean square deviation of
0.05~\cm; points
between dissociation and 75\,000 \cm\ are reproduced with
the average accuracy of a few wavenumbers. 
The new PES75K+ potential combines BOPES75K with  
adiabatic, relativistic  and quantum electrodynamics (QED) 
surfaces to provide the most accurate representation 
of the \hp\ global potential to date, overcoming the 
limitations  on previous high accuracy H$_3^+$ PESs near and above dissociation.
PES75K+ can be used 
to provide predictions of bound rovibrational energy levels
with an accuracy of approaching 0.1~\cm. Calculation of rovibrational energy levels
within PES75K+ suggests that the non-adiabatic
correction remains a limiting factor. 
The PES is also constructed to give the correct asymptotic limit making it suitable for
use in studies of the H$^+$\,+\,H$_2$ prototypical chemical reaction.
An improved dissociation energy  for H$_3^+$ is derived as $D_0\,=\,$35\,076\,$\pm\,2\,$cm$^{-1}$.


\end{abstract}

\bigskip

\section{Introduction}

The H$_3^+$ ion provides a benchmark system for two areas of science,
which, up to now, have remained unrelated: the high accuracy, \ai\
prediction of rotation-vibration spectra \cite{jt706} and reaction
dynamics \cite{14GoHoxx.H3+}. In fact, the two regimes are linked
through the near-dissociation spectrum of Carrington and co-workers
\cite{carrington:1982,carrington:1984,carrington:1986,carrington:1989,carrington:1993,00KeKiMc.H3+},
which provides a direct connection between spectroscopy and
dissociation dynamics. Theoretical studies to elucidate this spectrum
\cite{jt186}, as well as studies, which try to model ultra-low energy
H$^+$ + H$_2$ reactive and non-reactive
scattering~\cite{08CaGoRo.H3+,13HoScxx.H3+,15LaJaAo.H3+}, require
surfaces of spectroscopic accuracy to recover the full resonance
structure. 

A number of global potential energy surfaces (PESs) are available for
H$_3^+$ in its electronic ground state
\cite{95IcYoxx.H3+,jt202,jt247,aguado:2000,07ViAlVa.H3+,08VeLeAg.H3+,09BaCeJa.H3+,11BaPrVi.H3+,jt526,JAQ12:154307,MUK16:012050}.
Of these we particularly note the GLH3P PES of Pavanello \etal\
\cite{jt526}, which is based on \ai\ calculations of spectroscopic
accuracy \cite{jt512,jt535}, and the surface of Velilla \etal\
\cite{08VeLeAg.H3+}, which is based on less accurate \ai\ calculations
but whose full treatment of the long-range proton --
H$_2$ interactions is vital for the study of ultra-low energy
collisions \cite{15LaJaAo.H3+}.

The \ai\ calculations of Pavanello \etal\ \cite{jt526} 
used optimized explicitly correlated shifted Gaussian functions and
were performed for an extensive grid of  41 655 H$_3^+$ geometries. 
The calculations have an
absolute accuracy of about 0.01~\cm\ for the non-relativistic,
fixed-nuclei, electronic energy \cite{pa097}; Pavanello \etal's
analytical fit reproduces these points below dissociation with a root
mean square error of about 0.05 \cm.  However, certain asymptotic
configurations were omitted from this fit as their inclusion led to
significant deterioration in the accuracy at low energies. Therefore,
these points are not well represented by the analytical GLH3P yielding
unphysical features as shown in the supplemental material of the
original paper \cite{jt526}.  Such unphysical features also appear in other
global H$_3^+$ PESs \cite{jt247}. Importantly, we note
that the PES  of Velilla \etal\
\cite{08VeLeAg.H3+} does not show any unphysical behaviour.

The work of  Pavanello \etal\ \cite{jt526} also included an
adiabatic or diagonal Born-Oppenheimer correction, but did not provide a global fitted  surface for this effect. 
However, to make a PES spectroscopically accurate,  global adiabatic,
relativistic and QED (quantum electrodynamics) correction surfaces should also be included.
Such surfaces are available \cite{jt706}. 
In the present work, we report a significantly improved fit to the \ai\ points of Pavanello \etal\ \cite{jt526} in the high-energy region,
so that a very accurate global PES is obtained, which includes BO
energies, QED, relativistic and adiabatic corrections. This surface, which we call PES75K+, is suitable for both
the computation of predissociation energies and line positions, as well as for accurate reactive and non-reactive
scattering calculations. 

The present  paper is organized as follows: In section~\ref{pot_struct} we describe
the functional form of the improved global \hp\ PES. 
The global PES obtained as an analytical form fitted to the accurate \ai\
points is analyzed in section~\ref{res}, while section~\ref{res_nuc} contains a
comparison of the rovibrational energy level calculations using the present surface and 
our previous one. Section~\ref{conclusions} concludes our paper. 

\section{Structure of the new global potential}\label{pot_struct}

\subsection{Born-Oppenheimer part of the potential}\label{BO_part}
\begin{figure}[h!]
\center{\includegraphics{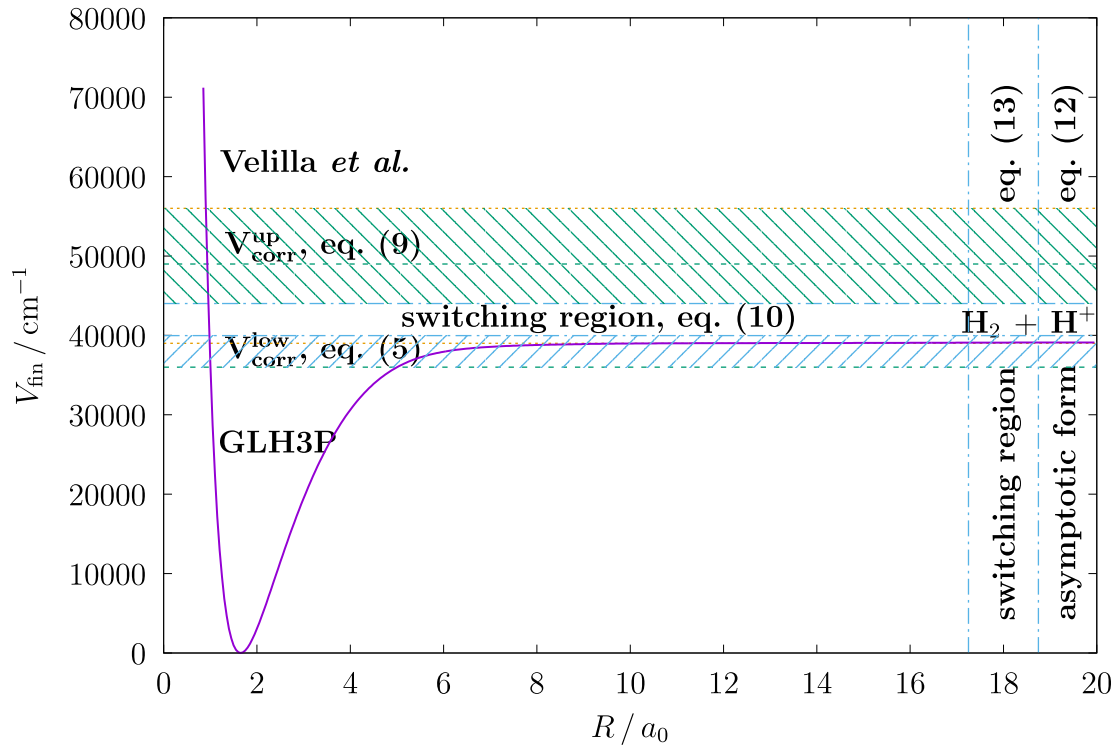}}
\caption{Illustration of the structure of new BOPES75K potential energy
surface. The curve corresponds to isosceles configurations of H$_3^+$ with the 
third bond length fixed at its equilibrium value $1.65\,a_0$.
 Dashed lines mark the borders of regions, where the GLH3P PES (the green line) and
 the potential of Velilla \etal\ (the yellow line) were corrected, and 
the blue dashed lines denote the borders of switching regions. The shaded domains correspond 
to the parts of corrected potentials, which don't fall into the energy switching region.}
\label{Schem_PES75K}
\end{figure}

Initially we attempted to fit the \ai\ data of Pavanello \etal\
\cite{jt526} using a single functional form. However, our attempts
failed to give a satisfactory fit.  We there adopted an alternative
approach where the BO part of the new \emph{ab initio} high-accuracy
global potential energy surface for the H$_3^+$ molecule consists of
three independent parts: low and high energy short-range potentials
joined together smoothly at 42\,000~cm$^{-1}$ using the energy
switching approach of Varandas~\cite{VAR96:3524}, and an analytic
long-range part, which gives the correct asymptotic behaviour. The
lower part of the short-range potential, $V_{\rm low}$, consists of
the three-body terms from the GLH3P potential obtained by Pavanello
\emph{et al.} \cite{jt526} with a correction in the energy region
where the GLH3P surface is not accurate.  The upper part, $V_{\rm
  up}$, is an analogous adaptation of the PES of Velilla \etal\
\cite{08VeLeAg.H3+}. The long-range potential is based on the analytic
form used by Velilla \etal\ \cite{08VeLeAg.H3+}.  We call the
resulting PES BOPES75K. Figure \ref{Schem_PES75K} illustrates the
structure of BOPES75K.

The BOPES75K potential extensively utilizes switching functions, 
\begin{equation}
f(t-t_0,\alpha) = \frac{1}{1+e^{-2\alpha \Delta t}} = \frac{1}{2}\left[1+\tanh{(\alpha 
\Delta t)}\right],
\end{equation}
where $\alpha$ is a parameter that controls the sharpness of the switch (large $\alpha$ is 
a fast switch and small $\alpha$ is a slow switch), $t_0$ is the switching point in variable $t$ 
and $\Delta t = t - t_0$ is the distance from the position of the switch. 
However, it is often easier to set a width for the switching zone, $b$, 
so that at the edges of the region $(t_0 - \frac{b}{2}, t_0 + \frac{b}{2})$ 
the switching function would reach its asymptotic values (0 or 1, correspondingly) 
with an accuracy of $\delta$:  
$$f(t_0 + \frac{b}{2}) = 1 - \frac{\delta}{100},$$
$$f(t_0 - \frac{b}{2}) = \frac{\delta}{100},$$
where $\delta$ is expressed as a percentage. So, the sharpness of the switch is fully defined by the width of the switching zone and the accuracy $\delta$ and can be expressed as

\begin{equation}\label{alpha}
\alpha = \ln\left(\frac{100}{\delta} - 1\right) \left/ b \right..
\end{equation}
We will use the parameters $t_0$, $b$ and $\delta$ later as variables in some optimization procedures.

A product of two switching functions can be used to smoothly turn on
and off a term within a finite region by using oppositely signed
switching parameters. A similar approach with multiple switchings was applied to the NO$_2$
molecule by Varandas in \cite{03Varand.NO2}. 
In this work we use switching functions based
directly on the coordinates and on the value of the potential energy
at a given geometry. In the latter case it is necessary to map from
the given coordinates to a value of the PES at that point in order to
evaluate the switching function. Below $V_d$ denotes the value of this
``distributing'' potential; where necessary the global H$_3^+$
potential of Velilla \etal\ \cite{08VeLeAg.H3+} was used to evaluate
$V_d$.

The correction to $V_{\rm low}$ was made by first approximating
the differences between the GLH3P energy values and 
the corresponding \emph{ab initio} energy values at 2414 \emph{ab initio} geometries 
in the range 35\,000 to 42\,000~cm$^{-1}$ using a polynomial 

\begin{equation}\label{pol}
V_{\rm poly}^{X} = \sum_{nmk}^N{V_{nmk}^{X}\,S_a^n\,S_e^{2m + 3k} \cos(3k\varphi)},\,n + 2m + 3k \leq N,
\end{equation}
where the superscript $X$ on $V_{\rm poly}^{X}$ and $V_{nmk}^{X}$
is used to distinguish the various expansions,  in this case $X={\rm low}$ for the low-energy short-range PES
and $N = 31$ coefficients with a maximum degree of 7.  This expression
uses the following symmetrized coordinates:
\begin{equation}\label{S_coords}
\begin{aligned}
S_a &= (\Delta R_{12} + \Delta R_{23} + \Delta R_{31})/\sqrt{3}, \\
S_x &= (2\Delta R_{12} - \Delta R_{23} - \Delta R_{31})/\sqrt{6} &= S_e \cos{\varphi},\\
S_y &= (\Delta R_{23} - \Delta R_{31})/\sqrt{2} &= S_e \sin{\varphi},
\end{aligned}
\end{equation}
where $\Delta R_{ik} = R_{ik} - R_e$ is the displacement from the equilibrium value 
of $R_e = 1.65\,a_0$ in the bond
length between the $i$-th and $k$-th protons in H$_3^+$ and the angle $\varphi$ from eq.~\eqref{pol} can be obtained from the latter equations as $\varphi = \arctan\left(\frac{S_y}{S_x}\right)$.
 This analytical form
reflects correctly the $D_{3h}$ symmetry of the H$_3^+$ molecular
system. The standard deviation of the approximated energy differences
from the numerical values of the polynomial
$V_{\text{poly}}^{\text{low}}$ in the corresponding \emph{ab initio}
geometries is about 2.3~cm$^{-1}$.

We added the possibility of smoothly switching on and off this polynomial term 
by multiplying it with switching factors $f_{low}$ and $f_{up}$:
\begin{equation}\label{Vcl}
V_{\rm corr}^{\rm low} = V_{\rm poly}^{\rm low}  f_{\rm low}  f_{\rm up},
\end{equation}
where
\begin{equation}\label{corr_terms}
\begin{aligned}
f_{\rm low} &= \frac{1}{1 + e^{-2\alpha_l(V_d - E_l)}}~, \\
f_{\rm up} &= \frac{1}{1 + e^{2\alpha_u(V_d - E_u)}}~, \\
\alpha_{l,u} &= \ln\left(\frac{100}{\delta} - 1\right) \left/ b_{l,u}\right..
\end{aligned}
\end{equation}

Here and elsewhere $\delta =0.1\,\%$, and $b_u$, $b_l$, $E_u$ and $E_l$
are the adjustable nonlinear parameters of the fit. The lower
part of the potential is thus represented as
\begin{equation}\label{Vlow}
V_{\rm low} = V_{\text{GLH3P}} - V_{\rm corr}^{\rm low}. \\
\end{equation}
The upper part of our new global H$_3^+$ BO PES is based
on the potential of Velilla \emph{et al.}~\cite{08VeLeAg.H3+}, denoted $V_{\text{Vel.}}$, 
as the GLH3P potential is not reliable in this energy region. It is corrected in the same
way as the lower part: 
\begin{equation}\label{Vup}
V_{\rm up} = V_{\text{Vel.}} - V_{\rm corr}^{\rm up}. 
\end{equation}
The correction term is also a polynomial $V_{\rm poly}^{\rm up}$ given
by eq.~\eqref{pol} with $N = 30$ coefficients $V_{nmk}^{\rm up}$ and a maximum
degree of 7. This function approximates a set of 4582 differences
between the $V_{\text{Vel.}}$ values and the accurate \emph{ab initio}
energies in the region from 40\,000~cm$^{-1}$ to 55\,000~cm$^{-1}$
with a standard deviation of about 4.6~cm$^{-1}$.
Again the function switches on and off
\begin{equation}\label{Vcu}
V_{\rm corr}^{\rm up} = V_{\rm poly}^{\rm up}  f_{\rm low}  f_{\rm up},
\end{equation}
where $f_{low}$ and $f_{up}$ are also given by eq.~\eqref{corr_terms}, just like
the ones used to describe $V_{\rm corr}^{\rm low}$, but with their own values of
adjustable nonlinear parameters. 

The two parts were then merged using the energy switching
scheme~\cite{VAR96:3524} to yield a composite potential
\begin{equation}\label{V_comp}
V_{\text{comp}} = \varphi_{low} V_{low} + \varphi_{up} V_{up},
\end{equation}
where 
\begin{equation}\label{Varandas_pars}
\begin{aligned}
\varphi_{low} &= \frac12(1 + \tanh{(-\gamma(V_{up} - E_0))}, \\
\varphi_{up} &= \frac12(1 + \tanh{(\gamma(V_{up} - E_0))}, \\
\gamma &= \tilde{g}_0 + \tilde{g}_1 (V_{up} - E_0)^2,
\end{aligned}
\end{equation}
and the coefficients $\tilde{g}_{0,1} = g_{0,1} / (100k^2),$ where $k =
219474.631\,\text{cm}^{-1}E_h^{-1}$ is a well-known conversion factor of energy
from $E_h$ to cm$^{-1}$ units.   

The leading term in the long-range behaviour of 
the H$_3^+$ potential at its first dissociation limit into H$_2\,+\,$H$^+$ 
is correctly represented in the GLH3P potential, as it is derived
from the analytical potentials of Viegas \emph{et al.}~\cite{07ViAlVa.H3+},
and therefore also in $V_{\text{comp}}$. However, the full, angularly-dependent asymptotic behaviour
is given by the multipole expansion~\cite{GEN75:90,08VeLeAg.H3+}
\begin{equation}\label{V_longrange}
V  - D_e \underset{R \to \infty}{\approx} \frac{Q_2(r)P_2(\cos \theta)}{R^3} 
- \frac{\frac{1}{2}a_0(r) + \frac{1}{3}\left[\alpha_{\parallel}(r) - \alpha_{\bot}(r)\right] P_2(\cos \theta)}{R^4}
+ \cdots + V_{H_2}(r),
\end{equation}
%
%
where $r$ is the internuclear separation of the diatomic (taken as the shortest distance between
two of the nuclei), $R$ is the distance between the midpoint between these two nuclei
and the third nucleus, and $\theta$ is the angle between $r$ and $R$.
In the case where two nearest nuclei are the same, these coordinates are standard Jacobi coordinates.
%
We enforce this behaviour in the asymptotic region by explicitly joining our BO surface defined 
in eq.~\eqref{V_comp} to the one due to Velilla \emph{et al.}, where this form is implemented. 
This is done at a certain distance $R_0$ using the switching procedure
\begin{align}
V_{fin} &= f_{short} V_{\text{comp.}} + f_{long}  V_{\text{Vel.}},
\end{align}
where
\begin{align}
f_{long} &= \frac{1}{1 + e^{-2\alpha(R - R_0)}}, \\
f_{short} &= \frac{1}{1 + e^{2\alpha(R - R_0)}}
\end{align}
and $\alpha$ is given by eq.~\eqref{alpha}.
R$_0$ and $b$ are two further nonlinear parameters in our final BO PES $V_{fin}$, 
and $\delta$ has the same value as earlier, see eq.~\eqref{corr_terms}.
This switching also ensures that the H$_2$ diatomic potential is correctly reproduced by
the $V_{fin}$ potential as $R \rightarrow \infty$.

Thus, the final form of our global BO PES contains 61 linear and 15 
nonlinear adjustable parameters of the fit,  including the parameters $E_0$, $g_0$ 
and $g_1$ for the energy switching~\cite{VAR96:3524}.
The linear parameters were determined by least-squares fitting; the nonlinear ones were adjusted manually by a 
trial-and-error procedure. Their final values are summarized in Table~\ref{tab_coefs} and Table~\ref{tab_coefs_pol}. 

\begin{table}[h]
\begin{center}
\caption{Adjustable nonlinear parameters used to define our 
new global \emph{ab initio} PES for the H$_3^+$ molecule together 
with the number $N_{\text{coef}}$ of linear polynomial parameters; 
all polynomials have a maximum degree of 7.}
\footnotesize
~\\
\begin{tabular}{lccrrccccccc}
\hline\hline
Part of the PES & $N_{\text{coef}}$ & $\delta$ & $b_u$/cm$^{-1}$ & $b_l$/cm$^{-1}$ & $E_u$/cm$^{-1}$ & $E_l$/cm$^{-1}$ & $E_0$/cm$^{-1}$ & $g_0$ & $g_1$ & R$_0$/$a_0$ & $b$/$a_0$\\
\hline
Lower (GLH3P-like)                 & $31$ & $0.1$ & $300$ & $100$ & $48\,000$ & $37\,000$ & \raisebox{-1.5ex}[0cm][0cm]{$42\,000$} & \raisebox{-1.5ex}[0cm][0cm]{$1.71$} & \raisebox{-1.5ex}[0cm][0cm]{$2000$} & \raisebox{-1.5ex}[0cm][0cm]{$18$} & \raisebox{-1.5ex}[0cm][0cm]{$1$}\\
Upper (Velilla \emph{et al.}-like) & $30$ & $0.1$ & $1400$ & $80$ & $55\,000$ & $40\,000$ & & & & & \\
\hline\hline
\end{tabular}
\label{tab_coefs}
\normalsize
\end{center}
\end{table}    

\begin{table}[ht!]
\begin{center}
\caption{Coefficients of the polynomials $V_{\text{poly}}^{\text{low}}$ and $V_{\text{poly}}^{\text{up}}$ 
given by eq.~\eqref{pol} for the correction terms to the lower and the upper parts of the new BOPES75K potential, correspondingly.}
\small
~\\
\begin{tabular}{cccrr}
\hline\hline
$n$ & $2m$ & $3k$ & $V_{nmk}^{\rm low}$ & $V_{nmk}^{\rm up}$ \\ 
\hline
0 & 0 & 0 & -0.75143071 & 28.89617715 \\
1 & 0 & 0 & -1.76440106 & -10.58157805 \\
2 & 0 & 0 & -0.09859390 & 2.87604608 \\
0 & 2 & 0 & 0.99200748 & 2.08628333 \\
3 & 0 & 0 & 1.48567895 & -0.62636049 \\
1 & 2 & 0 & -4.46772673 & -0.51256565 \\
0 & 0 & 3 & -3.90622308 & -1.15962111 \\
4 & 0 & 0 & -0.37926323 & 0.13476748 \\
2 & 2 & 0 & 1.53719611 & -0.08342574 \\
1 & 0 & 3 & 2.48855132 & 0.29409500 \\
0 & 4 & 0 & 1.37657184 & 0.10614684 \\
5 & 0 & 0 & -0.01544350 & -0.02831679 \\
3 & 2 & 0 & 0.07169542 & 0.08313826 \\
2 & 0 & 3 & -0.37054284 & -0.01860622 \\
1 & 4 & 0 & -0.68993265 & -0.10208867 \\
0 & 2 & 3 & -0.20471568 & -0.03815698 \\
6 & 0 & 0 & 0.00037056 & 0.00295364 \\
4 & 2 & 0 & 0.00167004 & -0.01213435 \\
3 & 0 & 3 & 0.00764265 & 0.00056655 \\
2 & 4 & 0 & -0.00024218 & 0.01681600 \\
1 & 2 & 3 & 0.02173168 & 0.00053141 \\
0 & 6 & 0 & 0.04686634 & -0.00595500 \\
0 & 0 & 6 & -0.00943583 & -0.00116108 \\
7 & 0 & 0 & 0.00077371 & -0.00010578 \\
5 & 2 & 0 & -0.00681003 & 0.00050616 \\
4 & 0 & 3 & -0.00285356 & -0.00000479 \\
3 & 4 & 0 & 0.01863322 & -0.00079326 \\
2 & 2 & 3 & 0.01297023 & 0.00002889 \\
1 & 6 & 0 & -0.01472313 & 0.00040228 \\
1 & 0 & 6 & 0.00114263 & 0.00004892 \\
0 & 4 & 3 & -0.01094351 &  \\
\hline\hline
\end{tabular}
\label{tab_coefs_pol}
\end{center}
\end{table} 

\subsection{Correction surfaces}\label{corrs}
It is necessary to also take into account adiabatic, relativistic and
QED corrections to the BO
approximation. The final form of our potential PES75K+ was obtained by
addition of these correction surfaces to the BOPES75K PES.

All correction surfaces consist of two parts: a polynomial given in the form of   eq.~\eqref{pol}, 
and an exponential damping function, which prevents unphysical behaviour of the corrections for
geometries with large internuclear distances.

\subsubsection{Relativistic and QED correction surfaces}
The polynomial parts $V_\text{poly}^\text{rel, QED}$ of relativistic and
QED correction surfaces have the analytical form of eq.~\eqref{pol}, but
different sets of adjustable parameters: $N=52$ for  the relativistic
correction, which correspond to maximum polynomial power of 10, and $N=44$
coefficients $V_{nmk}^\text{QED}$ for the QED correction,
with maximum polynomial power of 9. These sets of parameters were
obtained from two fits at points with energies below
38\,000~cm$^{-1}$: (i)~a fit of 3380 relativistic points by Bachorz
\emph{et al.}~\cite{Bachorz2009} to $V_\text{poly}^{rel}$, which are reproduced
with a standard deviation (rms) value of 0.008~cm$^{-1}$, and
(ii) a fit of 6413 QED points by Lodi~\emph{et al.}~\cite{jt581} to
$V_\text{poly}^\text{QED}$ with the rms deviation of 0.001~cm$^{-1}$.

Our estimates show that the overall effect of relativistic and QED
corrections on H$_3^+$ energy states even in the near-dissociation
region is only about 0.1~cm$^{-1}$, which is much smaller than the
corresponding BOPES75K accuracy of a few cm$^{-1}$, and we can
neglect this effect for states with large internuclear distances in
the region of the first dissociation limit and above.

We included the relativistic and the QED correction
surfaces using a combined polynomial form $V_\text{poly}^\text{rel + QED}$
given by eq.~\eqref{pol}, with maximum power of 10 and polynomial
coefficients $V_{nmk}^{\rm rel + QED} = V_{nmk}^{\rm rel} +
V_{nmk}^{\rm QED}$.  The combined polynomial correction
$V_\text{poly}^\text{rel + QED} = V_\text{poly}^\text{rel} +
V_\text{poly}^\text{QED}$ is then complemented by an exponentially
decreasing term, which smoothly switches the combined correction off
when its value becomes too large because of increasing internuclear
distances:

\begin{equation}\label{rel_qed_corr_exp_term}
\begin{aligned}
V_\text{rel + QED} &= V_\text{poly}^{\rm rel + QED} \cdot h, \\
h &= \frac{1}{1 + e^{2\alpha_c(|V_\text{poly}^{\rm rel + QED}| - V_0)}}, \\
\alpha_c &= \ln\left(\frac{100}{\delta} - 1\right) \left/ b_c\right.,
\end{aligned}
\end{equation}
where $\delta = 0.1$, $b_c = 3$~cm$^{-1}$, $V_0 = 6$~cm$^{-1}$. 
  
\subsubsection{Adiabatic correction surface}
The present adiabatic correction surface is a slightly modified
version of the one computed by us previously \cite{jt706}. Its polynomial part
$V_\text{poly}^\text{ad}$ has again the analytical form of eq.~\eqref{pol}, with a
set of 78 non-zero coefficients $V_{nmk}^{\rm ad}$ and a maximum
polynomial power of 12, but in this case transformed coordinates
$$\tilde{R}_{ij} = \left[1 - e^{-\beta(R_{ij}/R_e - 1)}\right] /
\beta$$ with $\beta = 1.3$ were used instead of the differences
$\Delta R_{ij}$. Parameters $V_{nmk}^{\rm ad}$ were obtained by fitting
5591 adiabatic points computed by Pavanello \emph{et al.}~\cite{jt526},
corresponding to energies up to 38\,000~cm$^{-1}$, to
$V_\text{poly}^{\rm ad}$. The resulting git gave an rms value of 0.116~cm$^{-1}$.

Unlike relativistic and QED corrections, the adiabatic one has an effect on
H$_3^+$ energy states in the near-dissociation region, which is comparable with
the corresponding BOPES75K accuracy. To extrapolate the adiabatic correction
surface to the region above 38\,000~cm$^{-1}$, we took an adiabatic point, which
is close to the dissociation limit, and considered its value $V_0^{\rm ad} =
-114.5\,$cm$^{-1}$ as a constant asymptote for H$_3^+$ molecular configurations
with two internuclear distances greater than $R_0^{\rm ad} = 10\,a_0$:

\begin{equation}\label{ad_corr_exp_term}
\begin{aligned}
V_\text{ad} &= V_\text{poly}^{\rm ad} h_{\rm low} + V_0^{\rm ad} h_{\rm up}, \\
h_{\rm low} &= \frac{1}{1 + e^{2\alpha_c^\text{ad}(R - R_0^{\rm ad})}}, \\
h_{\rm up} &= \frac{1}{1 + e^{-2\alpha_c^\text{ad}(R - R_0^{\rm ad})}}, \\
\alpha_c^{\rm ad} &= \ln\left(\frac{100}{\delta} - 1\right) \left/ b_c^{\rm ad}\right.,
\end{aligned}
\end{equation}
where $\delta = 0.1$, $b_c^\text{ad} = 1$~cm$^{-1}$, and distance $R$ has the same meaning as earlier 
in eq.~\eqref{V_longrange}. 

The Fortran files with BOPES75K and global correction surfaces together with
files containing their polynomial constants are presented in the supplementary
material.

\section{Properties of the new global BO PES}\label{res}

In this work we used a set of 40\,537 \emph{ab initio} energies computed by
Pavanello \emph{et al.}~\cite{jt526}, which
span energies up to 75\,500~cm$^{-1}$ and are reproduced by the new potential
BOPES75K with a standard deviation value of 4.9~cm$^{-1}$. 786 points (about 2\% from
the total set of 41\,323 geometries) with energy values from 37\,090~cm$^{-1}$
to 75\,380~cm$^{-1}$ were excluded from our calculations, because their
inclusion significantly deteriorates the accuracy. About 20\% of them have two
large (greater than 7\,$a_0$) internuclear distances and energies up to
42\,000~cm$^{-1}$, and the others correspond to energies above 53\,000~cm$^{-1}$
and have comparatively small bond lengths. 
The excluded points are nevertheless represented reasonably well by BOPES75K, with
a standard deviation of 40.4~cm$^{-1}$ and a 
maximum deviation of about 170~cm$^{-1}$.

\begin{figure}[h]
\center{\includegraphics{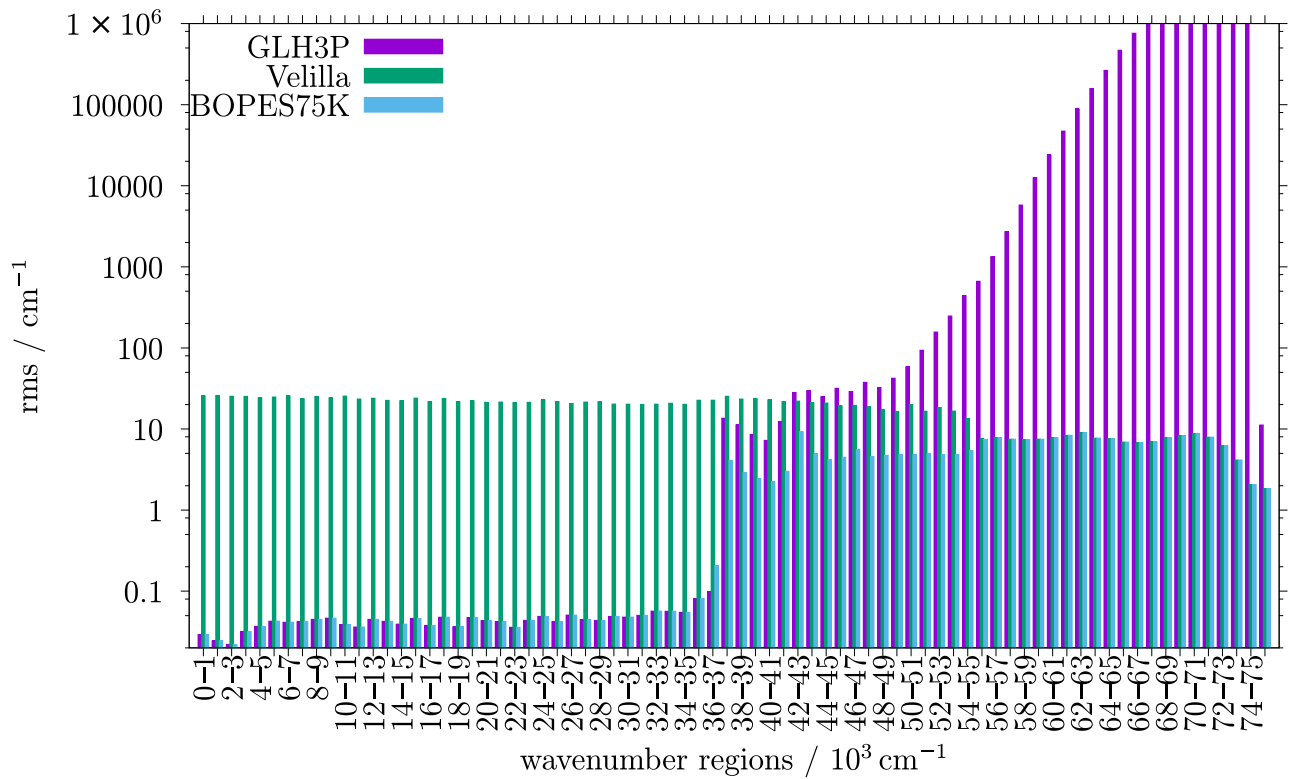}}
\caption{Standard deviations, in cm$^{-1}$, with which various potentials reproduce the
high accuracy \emph{ab initio} points computed by
Pavanello \emph{et al.}~\cite{jt526}: GLH3P potential \cite{jt526}, the PES by Velilla \emph{et al.}
\cite{08VeLeAg.H3+} and the new BOPES75K potential. Energy ranges all have equal widths of 1000\,cm$^{-1}$.}
\label{hist_sd}
\end{figure}

Table~\ref{tab_res} and Figure~\ref{hist_sd} compare 
the GLH3P potential, the double many-body expansion (DMBE) potential of Viegas \emph{et al.} \cite{07ViAlVa.H3+}, 
whose function form is the basis of the GLH3P PES, the PES by Velilla \emph{et al.} and our new 
BOPES75K for different energy ranges.
It is clear that our new BO PES retains the accuracy of the GLH3P fit at
low energy while greatly improving its behaviour near and above dissociation.

\begin{table}[ht!]
\begin{center}
\caption{ Standard deviations, in cm$^{-1}$, of PES energy minus
 the \emph{ab initio} value  for the GLH3P potential \cite{jt526}, the DMBE potential of 
Viegas \emph{et al.} \cite{07ViAlVa.H3+}, the PES by Velilla \emph{et al.}
\cite{08VeLeAg.H3+} and the new BOPES75K potential for different energy ranges.
The zero energy for all calculations was taken from the GLH3P potential. $N(E)$ denotes 
the number of \ai\ points falling in the given energy range.
The rms values indicated by a star* correspond to energy 
regions that are not covered by the GLH3P surface.}
\small
~\\
\begin{tabular}{crrrrr}
\hline\hline
Energy range/cm$^{-1}$ & $N(E)$ & DMBE & GLH3P & 
Velilla \emph{et al.} & BOPES75K \\ 
\hline
0---35\,000       & 5422  & 30.56  &   0.0447  & 22.34 & 0.0447 \\
0---37\,000       & 6005  & 60.42  &   0.0512  & 22.37 & 0.0666 \\
37\,000---42\,000 & 1841  & 151.30 &  11.18  & 23.68 & 3.070 \\
42\,000---45\,000 & 934   & 121.90 &  27.65  & 21.38 & 5.611 \\
45\,000---50\,000 & 1503  & 120.71 &  34.99* & 18.32 & 4.861 \\
50\,000---55\,000 & 1690  & 111.27 & 246.34* & 17.10 & 4.981 \\
53\,000---75\,500 & 29260 & 84.12  &  1.4E+07* &  5.872 & 5.443 \\
\hline\hline
\end{tabular}
\label{tab_res}
\end{center}
\end{table} 

\begin{figure}[h!]
\center{\includegraphics{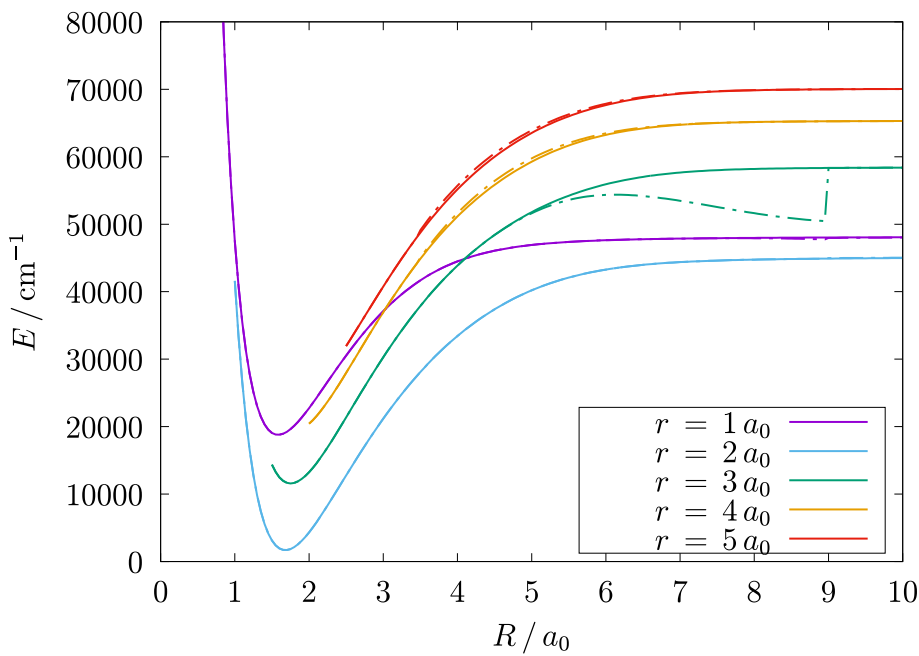}}
\caption{Comparison of some isosceles two-dimensional cuts of the new global
\emph{ab initio} BOPES75K for H$_3^+$ (solid lines) with the ones of the old
GLH3P potential (dashed lines). Here the bond lengths of the H$_3^+$ molecule
are $r_1 = r$, $r_2 = r_3 = R$.}
\label{c2v_cuts}
\end{figure}

Some two-dimensional cuts of our new global \emph{ab initio} BOPES75K for H$_3^+$ 
are pictured in Fig.~\ref{c2v_cuts}. They demonstrate the smooth behaviour of our new BO PES, 
 in contrast to the analogous  cuts through the GLH3P potential, which show serious
 unphysical features. 

\begin{figure}[h!]
\center{\resizebox{1.1\textwidth}{!}{\includegraphics{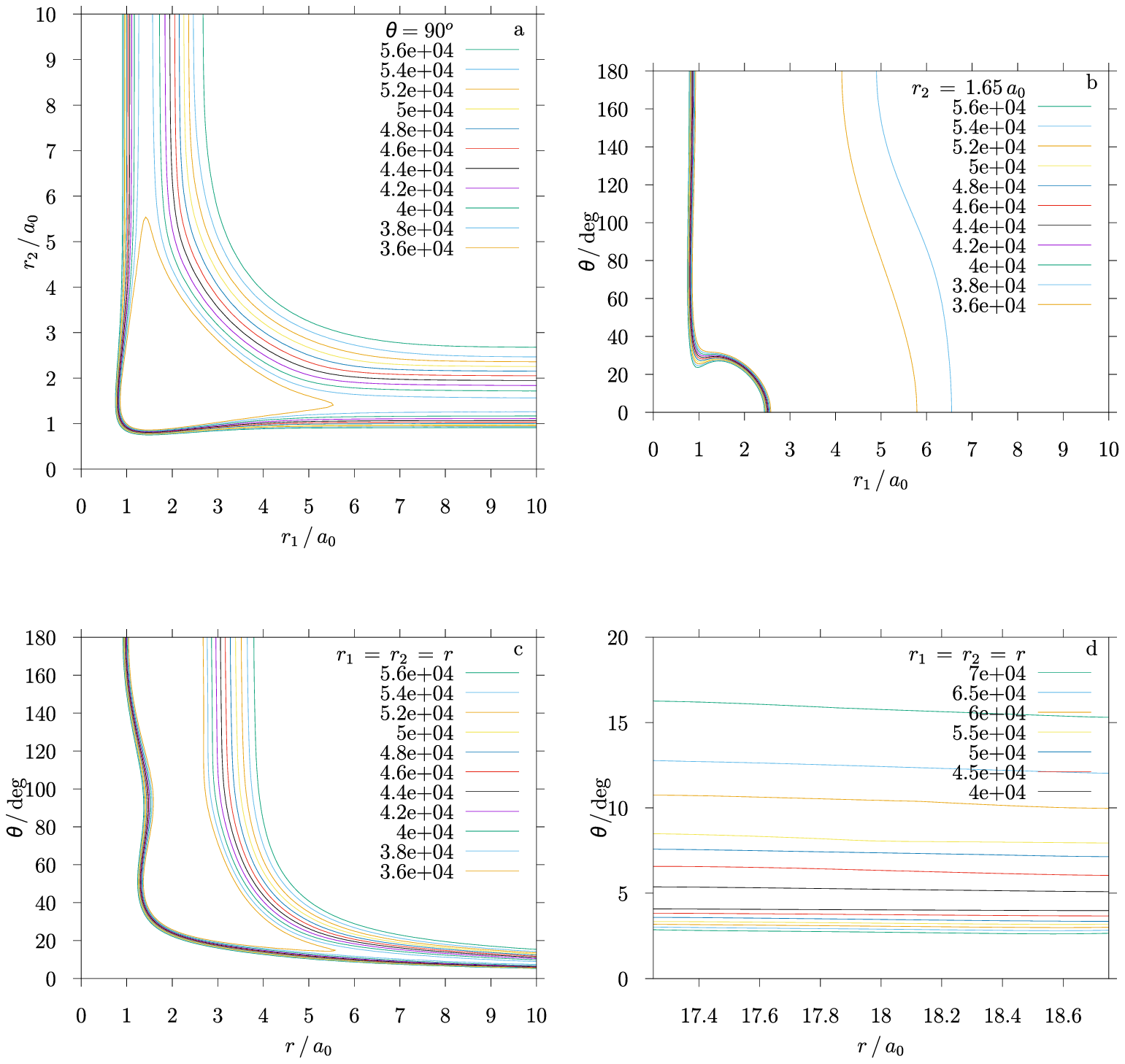}}}
\caption{Contour plots of BOPES75K in the
switching regions using H -- H bondlengths ($r_1,r_2$) and the angle between them ($\theta$): (a) configurations with 
$\theta=90^\circ$  in the energy switching
region from 36\,000 to 56\,000\,cm$^{-1}$; (b) configurations
with $r_2 =1.65$\,$a_0$, in the same
energy switching region; (c) isosceles configurations with $r_1 =
r_2 = r$, in the same energy switching region; (d) isosceles
configurations with $r_1 = r_2 = r$, for asymptotic switching region with $r$
varied from 17.25 to 18.75$\,a_0$. Energy values, which correspond to various
isolines, are indicated on plots in cm$^{-1}$.}
\label{c2v_isolines}
\end{figure}

Some two-dimensional cuts and contour plots of our new global \emph{ab initio}
BOPES75K for H$_3^+$ 
are shown in Fig.~\ref{c2v_cuts}, which compares with the
GLH3P potential,   and Fig.~\ref{c2v_isolines}, which illustrates the behaviour in switching regions.
The plots demonstrate the smooth behavior of our new BO PES; 
in contrast some of the analogous cuts through the
GLH3P potential show serious
 unphysical features.

The new BOPES75K can be used to give the dissociation limit into H$_3^+$
$\rightarrow$ H$_2\,+\,$H$^+$ $D_e\,=\,$37\,195.3\,cm$^{-1}$.  This is
due to the procedure of obtaining the new BO potential presented in
Sec.~\ref{pot_struct}. For H$_2\,-\,$H$^+$ distances larger than $R =
19\,a_0$, the absolute values of differences between the BOPES75K and
the original potential by Velilla \emph{et al.} are less than
0.005\,cm$^{-1}$. In particular, the PES of Velilla \emph{et al.}
reproduces a set of high-precision \emph{ab initio} points obtained by
them in \cite{08VeLeAg.H3+} with a standard deviation of only
1.80\,cm$^{-1}$ for H$_2\, - $H$^+$ distances $R$ larger than
10$\,a_0$ in the region of the first dissociation limit, and gives a
value of the BO dissociation energy $D_e\,=\,$37\,170\,cm$^{-1}$.

The BO value of dissociation energy $D_0$ obtained on the basis of our
new BOPES75K is $D_0\,=\,$35\,011.8\,cm$^{-1}$. In this calculation an
experimental value $E_0(\text{H}_2)\,=\,2179.3\,\pm\,0.1$\,cm$^{-1}$
for H$_2$ vibrational zero-point energy (ZPE) was used
\cite{07Irikur.gen}.  The vibrational ZPE for H$_3^+$
$E_0(\text{H}_3^+)\,=\,4362.76$\,cm$^{-1}$ was obtained from the large
basis set calculation performed in Sec.~\ref{res_nuc} which is enough
to converge the ZPE within 0.0001\,cm$^{-1}$. For comparison with experiment
a further 64.21\,cm$^{-1}$ \cite{jt318} must be added to the ZPE as the Pauli Principle
means that the lowest allowed state of H$_3^+$ has is the $J=1,K=1$.
The
influence of disregarding non-BO effects is mainly conditioned by the
adiabatic correction: the error from neglecting relativistic and QED
effects is probably less than 1\,cm$^{-1}$ (as in calculations
for the H$_2$ molecule \cite{16PuKoCz.H2}), and the value of nonadiabatic correction is
negligible at the dissociation threshold. Our results for the adiabatic
correction at equilibrium give a value about 115.1\,cm$^{-1}$,
whereas its value near dissociation was taken equal to be
114.5\,cm$^{-1}$ (see Sec. \ref{corrs}). Thus, our estimate of the uncertainty
in our calculated dissociation energy  is about 2\,cm$^{-1}$, and
$D_0\,=\,$35\,076\,$\pm\,2\,$cm$^{-1}\,=\,4.3489\,\pm\,0.0002\,$eV.
This value is a little higher than the best previously-available
theoretical result due to Lie \& Frye \cite{lf92} of
$4.337\,\pm\,0.002\,$eV. However, it remains lower than the best
available experiment estimate of $D_0$\,=\,$4.381\,\pm\,0.021\,$eV due
to Cosby and Helm \cite{cosby:1988}.  We believe our value is the best
available estimate of the dissociation energy of H$_3^+$.



\section{Nuclear motion calculations}\label{res_nuc}

In order to test the new PES75K+ surface, calculations 
were performed using the {\sc DVR3D} variational nuclear motion program suite~\citep{jt338}. 
The calculations  were performed in Jacobi coordinates, and 
the discrete variable representation  (DVR) grids were based on spherical
oscillator functions  for both the atom -- diatom
coordinate, $R$, and the diatomic  coordinate, $r$, defined by the 
parameters $\alpha = 0.0$ and
$\omega_e = 0.07$ atomic units \cite{jt23}. A DVR in (associated) Legendre
functions was used for the angular coordinate, $\theta$. The grids contained
$60$, $58$, and $68$ points for $R$, $r$ and $\theta$ coordinates, respectively.  The
final diagonalized matrices for the vibrational problem had a dimension
of $20\,000$. 

We compute vibrational energy levels, i.e. with total angular momentum
$J = 0$, using the new BOPES75K and the GLH3P PES for two cases: (i) for
the surfaces in BO approximation using nuclear masses only -- up to
60\,000~cm$^{-1}$, and (ii) for these PESs augmented by adiabatic,
relativistic and quantum electrodynamics correction
surfaces, and allowing for non-adiabatic effects by using
different effective vibrational and rotational masses
as was suggested by Moss \cite{96Moss}. Calculations were performed up to
the first dissociation limit of H$_3^+$ at about 37\,200~cm$^{-1}$, and
a bit above. This non-adiabatic model has been shown to give highly
accurate predictions of H$_3^+$ spectra \cite{jt236}. In the latter
calculation set with BOPES75K the new global correction surfaces with
exponential ``tails'' were used, and the vibrational mass was to
1.007517~Da -- a value intermediate between nuclear and atomic masses,
which is the optimal one for the issue of the most accurate prediction
of experimental vibrational band origins (see Table \ref{tab_exp}) and
was obtained manually by trial-end-error method. For calculations with
GLH3P adiabatic and
relativistic correction surfaces\cite{jt526}, and a QED correction surface \cite{jt581}, which do not
have global character, were used. In these calculations, the value of the vibrational mass
derived by Moss for H$_2^+$ \cite{96Moss} and used previously for H$_3^+$
\cite{jt526,jt236} was also employed. For calculations with $J > 0$ a rotational
mass is also needed; this was always set to the proton (nuclear) mass, as before \cite{jt526,jt236}.

The levels of \hp\, included in the calculations of Table~\ref{tab_diff_BO}
consist of two distinct sets -- bound and unbound levels.
The bound levels have a clear physical meaning, the unbound ones -- the levels
above the dissociation, are artefacts of the chosen basis set.
However, since the basis sets in both GLH3P and PES75K calculations
are the same, the discrepancies between these artefact levels display
the real difference between the resonances, which could be obtained
using the same PESs and, for example, a complex absorbing potential (CAP) 
\cite{93RiMexx.method,jt443}.

A comparison between the vibrational energies obtained in the 
PES75K and GLH3P calculations is performed in Tables~\ref{tab_diff_BO} 
and \ref{tab_diff_BO+corr}. One can see that there is a minor difference 
between the energies calculated on the basis of the two BO potentials up 
to the dissociation energy value, but significant 
differences (up to tens of cm$^{-1}$) appear in energy range from 
about 40\,000~cm$^{-1}$ and above. This is a direct consequence of fixing 
the unsmooth parts of the GLH3P PES.

Thus, Table~\ref{tab_diff_BO+corr} reflects only changes in the way 
non-BO effects are taken into account  as it covers only the energy
region up to 37\,000~cm$^{-1}$. The difference between the approaches
performed in the present work and previously in works by Pavanello
\emph{et al.} \cite{jt526} and Lodi \emph{et al.} \cite{jt581} becomes
significant for energies above 30\,000~cm$^{-1}$, i.e. in the energy region
where the old correction surfaces were not fitted accurately.

\begin{table}[ht!]
\begin{center}
\caption{Root mean square deviations (rms) and maximum absolute 
deviation (max) values in cm$^{-1}$ of the vibrational (with $J = 0$) 
Born-Oppenheimer energies obtained using the new global \emph{ab initio} 
potential BOPES75K from the ones calculated with the GLH3P PES for 
different energy ranges. $N_v(E)$ gives the number
of vibrational states given in each energy range.}
\small
~\\
\begin{tabular}{crrr}
\hline\hline
Energy range/cm$^{-1}$ & $N_v(E)$ & rms & 
max\\ 
\hline
0---5\,000        & 6     &   0.0000 & 0.0000 \\
5\,000---10\,000  & 21    &   0.0000 & 0.0001 \\
10\,000---15\,000 & 52    &   0.0001 & 0.0002 \\
15\,000---20\,000 & 110   &   0.0001 & 0.0007  \\
20\,000---25\,000 & 203   &   0.0004 & 0.0015  \\
25\,000---30\,000 & 331   &   0.0017 & 0.0067  \\
30\,000---35\,000 & 528   &   0.0070 & 0.0340  \\
35\,000---40\,000 & 776   &   0.0225 & 0.1924  \\
40\,000---45\,000 & 1043  &   0.2612 & 1.5575  \\
45\,000---50\,000 & 1310  &   1.8626 & 8.2676  \\
50\,000---55\,000 & 1583  &   3.8516 & 20.0573 \\
55\,000---60\,000 & 1832  &   4.4036 & 14.2542 \\
\hline\hline
\end{tabular}
\label{tab_diff_BO}
\end{center}
\end{table} 

\begin{table}[h!]
\begin{center}
\caption{Root mean square deviations (rms) and maximum absolute deviation (max) values in cm$^{-1}$ of vibrational 
(with $J = 0$) energy levels obtained using the new global \emph{ab initio} potential PES75K+ from the ones calculated 
with the GLH3P PES, for different energy ranges. Here adiabatic, relativistic, quantum electrodynamics 
and (partially) nonadiabatic effects were considered in following ways: for PES75K+ -- as  described above, 
for GLH3P PES -- as in Refs.~\cite{jt526} (adiabatic, relativistic and nonadiabatic parts) and \cite{jt581} (QED part).
$N_v(E)$ gives the number of vibrational states given in each energy range.}
\small
~\\
\begin{tabular}{cccr}
\hline\hline
Energy range/cm$^{-1}$ & $N_v(E)$ & rms & 
max\\ 
\hline
0---5\,000        & 6     &   0.0770 & 0.0983 \\
5\,000---10\,000  & 21    &   0.1485 & 0.1885 \\
10\,000---15\,000 & 52    &   0.2059 & 0.2524 \\
15\,000---20\,000 & 110   &   0.2410 & 0.2844  \\
20\,000---25\,000 & 203   &   0.2175 & 0.3291  \\
25\,000---30\,000 & 331   &   0.1281 & 0.4733  \\
30\,000---35\,000 & 528   &   1.6869 & 7.0708  \\
35\,000---38\,000 & 432   &   3.5309 &10.0955  \\
\hline\hline
\end{tabular}
\label{tab_diff_BO+corr}
\end{center}
\end{table} 

Finally, in Table \ref{tab_exp} a comparison of vibrational energy levels obtained using PES75K+ with adiabatic, relativistic, quantum electrodynamics, and (partially) nonadiabatic corrections with the available experimental data for states with $J = 0$ is performed. The standard deviation obtained in this comparison is about 0.12\,cm$^{-1}$, which is almost twice smaller than the value 0.21\,cm$^{-1}$ obtained with vibrational masses used by Moss \cite{96Moss}.

\begin{table}[h!]
\begin{center}
\caption{Comparison of vibrational band origins (energy levels with $J = 0$) obtained using the new global \emph{ab initio} potential PES75K+ with the available experimental data \cite{H3pMARVEL1}. Adiabatic, relativistic, quantum electrodynamics and (partially) nonadiabatic effects were considered in this set of calculations. The calculated levels with symmetry $E'$ were obtained as an arithmetic mean of (quasi-) degenerate even and odd pairs of DVR3D levels, which correspond to degenerate f-symmetry states. The experimental energies are supplemented by corresponding quantum numbers sets. Here ($\nu_1$, $\nu_2^{l_2}$) are vibrational quantum numbers, $K$ is an absolute value of the projection of $J$ on the C$_3$ axis and $G$ is an absolute value of quantum number $g = k - l_2$.}
\small
~\\
\begin{tabular}{lrrrcccccc}
\hline\hline
Symmetry & $E_{\text{exp}}$/cm$^{-1}$ & PES75K+/cm$^{-1}$ & Obs. - calc./cm$^{-1}$ & $K$ & $G$ & $\nu_1$ & $\nu_2$ & $l_2$\\ 
\hline
$E'$   & 2521.411  & 2521.307  & 0.104 & 0 & 1 & 0 & 1 & 1 \\
$E'$   & 4998.049  & 4997.912  & 0.137 & 0 & 2 & 0 & 2 & 2 \\
$E'$   & 5554.060  & 5554.223  &-0.163 & 0 & 1 & 1 & 1 & 1 \\
$A_2'$ & 7492.912  & 7492.807  & 0.106 & 0 & 3 & 0 & 3 & 3 \\
$E'$   & 9113.080  & 9113.077  & 0.003 & 0 & 2 & 0 & 4 & 2 \\
$E'$   & 10645.380 & 10645.338 & 0.042 & 0 & 2 & 2 & 2 & 2 \\
$E'$   & 10862.910 & 10862.780 & 0.130 & 0 & 1 & 0 & 5 & 1 \\
$E'$   & 11323.100 & 11323.145 &-0.045 & 0 & 1 & 3 & 1 & 1 \\
$E'$   & 11658.400 & 11658.344 & 0.056 & 0 & 5 & 0 & 5 & 5 \\
$E'$   & 12303.370 & 12303.376 &-0.006 & 0 & 1 & 2 & 3 & 1 \\
$E'$   & 12477.380 & 12477.432 &-0.052 & 0 & 2 & 0 & 6 & 2 \\
$E'$   & 13702.380 & 13702.676 &-0.296 & 0 & 1 & 0 & 7 & 1 \\
$E'$   & 15122.810 & 15122.725 & 0.085 & 0 & 2 & 0 & 8 & 2 \\
\hline\hline
\end{tabular}
\label{tab_exp}
\end{center}
\end{table} 

\section{Conclusions}
\label{conclusions}

We present a modified global \hp\ PES in both a simple BO form and as
an augmented BO plus relativistic, QED and adiabatic corrections PES.
The \ai\ points used for this representation \cite{jt512} are
extremely accurate (accurate to an absolute energy of about 0.01\,\cm)
and the only problem with the previous fit to these
points  was that analytical
representation of some configurations with energy values above
37\,000\,\cm\ was poor leading them to be excluded from the fit. The task of
representing all the \ai\ points with the intrinsic accuracy of \ai\
calculations proved to be difficult, and the improved PESs presented
here still do not recover the full accuracy of the underlying \ai\ electronic structure points.
Our new representation corrects the shortcomings of the previous
incarnation~\cite{jt526} and displays the correct asymptotic
behaviour. 

Our new PES is suitable for a variety of calculations.  Firstly, scattering
calculations for collisions between protons and diatomic hydrogen
molecules, particularly at ultra-low energies where such collisions
are very sensitive to the details of the underlying PES. Secondly, the
study of asymptotic vibrational states for the H$_3^+$ system \cite{jt358}. 
Thirdly, for accurate calculation of \hp\ resonance states,
 which should finally lead to the interpretation of the famous,
indeed notorious, Carrington-Kennedy predissociation
spectrum~\cite{CB82,CK84}.

\section*{Acknowledgements}
This work was partially supported by the Russian Fund for Fundamental 
Studies as part of the research project \# 18-02-00705 and by the European Union's Horizon
2020 research and innovation programme under the Marie
Sklodowska-Curie grant through grant number 701962.
We thank Eryn Spinlove and Attila Cs\'as\'ar for comments on our
potential energy surface.


\begin{thebibliography}{46}
\providecommand{\url}[1]{\texttt{#1}}
\providecommand{\urlprefix}{URL }
\markboth{Taylor \& Francis and I.T. Consultant}{Molecular Physics}

\bibitem{jt706}
J. Tennyson, O.L. Polyansky, N.F. Zobov, A. Alijah and A.G. Cs\'asz\'ar,  J.
  Phys. B: At. Mol. Opt. Phys.  \textbf{50}, 232001 (2017).

\bibitem{14GoHoxx.H3+}
T. Gonzalez-Lezana and P. Honvault,  Intern. Rev. Phys. Chem.  \textbf{{33}},
  371 ({2014}).

\bibitem{carrington:1982}
A. Carrington, J. Buttenshaw and R. Kennedy,  Mol. Phys.  \textbf{45}, 753
  (1982).

\bibitem{carrington:1984}
A. Carrington and R.A. Kennedy,  J. Chem. Phys.  \textbf{81}, 91 (1984).

\bibitem{carrington:1986}
A. Carrington,  J. Chem. Soc. Farad. T.2  \textbf{82}, 1089 (1986).

\bibitem{carrington:1989}
A. Carrington and I.R. Mc{N}ab,  Acc. Chem. Res.  \textbf{22}, 218 (1989).

\bibitem{carrington:1993}
A. Carrington, I.R. Mc{N}ab and Y.D. West,  J. Chem. Phys.  \textbf{98}, 1073
  (1993).

\bibitem{00KeKiMc.H3+}
F. Kemp, C.E. Kirk and I.R. McNab,  Phil. Trans. A  \textbf{358} (1774), 2403
  (2000).

\bibitem{jt186}
N.F. Zobov, O.L. Polyansky, C.R. {Le Sueur} and J. Tennyson,  Chem. Phys. Lett.
   \textbf{260}, 381 (1996).

\bibitem{08CaGoRo.H3+}
E. Carmona-Novillo, T. Gonzalez-Lezana, O. Roncero, P. Honvault, J.M. Launay,
  N. Bulut, F.J. Aoiz, L. Banares, A. Trottier and E. Wrede,  J. Chem. Phys.
  \textbf{{128}}, 014304 ({2008}).

\bibitem{13HoScxx.H3+}
P. Honvault and Y. Scribano,  J. Phys. Chem. A  \textbf{{117}}, 9778 (1997).

\bibitem{15LaJaAo.H3+}
M. Lara, P.G. Jambrina, F.J. Aoiz and J.M. Launay,  J. Chem. Phys.
  \textbf{{143}}, 204305 ({2015}).

\bibitem{95IcYoxx.H3+}
A. Ichihara and K. Yokoyama,  J. Chem. Phys.  \textbf{{103}}, 2109 ({1995}).

\bibitem{jt202}
R. Prosmiti, O.L. Polyansky and J. Tennyson,  Chem. Phys. Lett.  \textbf{273},
  107 (1997).

\bibitem{jt247}
O.L. Polyansky, R. Prosmiti, W. Klopper and J. Tennyson,  Mol. Phys.
  \textbf{98}, 261 (2000).

\bibitem{aguado:2000}
A. Aguado, O. Roncero, C. Tablero, C. Sanz and M. Paniagua,  J. Chem. Phys.
  \textbf{112}, 1240 (2000).

\bibitem{07ViAlVa.H3+}
L.P. Viegas, A. Alijah and A.J.C. Varandas,  J. Chem. Phys.  \textbf{{126}},
  074309 ({2007}).

\bibitem{08VeLeAg.H3+}
L. Velilla, B. Lepetit, A. Aguado, J.A. Beswick and M. Paniagua,  J. Chem.
  Phys.  \textbf{{129}}, 084307 ({2008}).

\bibitem{09BaCeJa.H3+}
R.A. Bachorz, W. Cencek, R. Jaquet and J. Komasa,  J. Chem. Phys.
  \textbf{131}, 024105 (2009).

\bibitem{11BaPrVi.H3+}
P. Barragan, R. Prosmiti, P. Villarreal and G. Delgado-Barrio,  Int.\ J.\
  Quant.\ Chem.  \textbf{111}, 368 (2011).

\bibitem{jt526}
M. Pavanello, L. Adamowicz, A. Alijah, N.F. Zobov, I.I. Mizus, O.L. Polyansky,
  J. Tennyson, T. Szidarovszky and A.G. {Cs\'asz\'ar},  J. Chem. Phys.
  \textbf{136}, 184303 (2012).

\bibitem{JAQ12:154307}
R. Jaquet and M.V. Khoma,  J. Chem. Phys.  \textbf{136}, 154307 (2012).

\bibitem{MUK16:012050}
B. Mukherjee, S. Mukherjee and S. Adhikari,  J. Phys. Conf.
  Ser.  \textbf{759} (1), 012050 (2016).

\bibitem{jt512}
M. Pavanello, L. Adamowicz, A. Alijah, N.F. Zobov, I.I. Mizus, O.L. Polyansky,
  J. Tennyson, T. Szidarovszky, A.G. Cs\'asz\'ar, M. Berg, A. Petrignani and A.
  Wolf,  Phys. Rev. Lett.  \textbf{108}, 023002 (2012).

\bibitem{jt535}
O.L. Polyansky, A. Alijah, N.F. Zobov, I.I. Mizus, R. Ovsyannikov, J. Tennyson,
  T. Szidarovszky and A.G. Cs\'asz\'ar,  Phil. Trans. Royal Soc. London A
  \textbf{370}, 5014 (2012).

\bibitem{pa097}
M. Pavanello and L. Adamowicz,  J. Chem. Phys.  \textbf{{130}}, 034104
  ({2009}).

\bibitem{VAR96:3524}
A.J.C. Varandas,  J.  Chem. Phys.  \textbf{105}, 3524
  (1996).

\bibitem{03Varand.NO2}
A.J.C. Varandas,  J. Chem. Phys.  \textbf{119}, 2596 (2003).

\bibitem{GEN75:90}
W.R. Gentry and C.F. Giese,  Phys. Rev. A  \textbf{11}, 90 (1975).

\bibitem{Bachorz2009}
R.A. Bachorz, W. Cencek, R. Jaquet and J. Komasa,  J. Chem. Phys.
  \textbf{131}, 024105 (2009).

\bibitem{jt581}
L. Lodi, O.L. Polyansky, A.A. J.~Tennyson and N.F. Zobov,  Phys. Rev. A
  \textbf{89}, 032505 (2014).

\bibitem{07Irikur.gen}
K.K. Irikura,  J.~Phys.\ Chem. ~Ref. ~Data  \textbf{36}, 389 (2007).

\bibitem{jt318}
J. Ramanlal, O.L. Polyansky and J. Tennyson,  Astron. Astrophys.  \textbf{406},
  383 (2003).

\bibitem{16PuKoCz.H2}
M. Puchalski, J. Komasa, P. Czachorowski and K. Pachucki,  Phys. Rev. Lett.
  \textbf{{117}}, 263002 ({2016}).

\bibitem{lf92}
G.C. Lie and D. Frye,  J. Chem. Phys.  \textbf{96}, 6784 (1992).

\bibitem{cosby:1988}
P.C. Cosby and H. Helm,  Chem. Phys. Lett.  \textbf{152}, 71 (1988).

\bibitem{jt338}
J. Tennyson, M.A. Kostin, P. Barletta, G.J. Harris, O.L. Polyansky, J. Ramanlal
  and N.F. Zobov,  Comput. Phys. Commun.  \textbf{163}, 85 (2004).

\bibitem{jt23}
J. Tennyson and B.T. Sutcliffe,  J. Mol. Spectrosc.  \textbf{101}, 71 (1983).

\bibitem{96Moss}
R.E. Moss,  Mol. Phys.  \textbf{{89}}, 195 ({1996}).

\bibitem{jt236}
O.L. Polyansky and J. Tennyson,  J. Chem. Phys.  \textbf{110}, 5056 (1999).

\bibitem{93RiMexx.method}
U.V. Riss and H.D. Meyer,  J.~Phys.~B: At.\ Mol.\ Opt.\ Phys.  \textbf{26},
  4503 (1993).

\bibitem{jt443}
B.C. Silva, P. Barletta, J.J. Munro and J. Tennyson,  J. Chem. Phys.
  \textbf{128}, 244312 (2008).

\bibitem{H3pMARVEL1}
T. Furtenbacher, T. Szidarovszky, E. M{\'a}tyus, C. F{\'a}bri and A.G.
  Cs{\'a}sz{\'a}r,  J. Chem. Theory Comput.  \textbf{9}, 5471 (2013).

\bibitem{jt358}
J.J. Munro, J. Ramanlal and J. Tennyson,  New J. Phys  \textbf{7}, 196 (2005).

\bibitem{CB82}
A. Carrington, J. Buttenshaw and R.A. Kenedy,  Mol. Phys.  \textbf{45}, 753
  (1982).

\bibitem{CK84}
A. Carrington and R.A. Kennedy,  J. Chem. Phys.  \textbf{81}, 91 (1984).

\end{thebibliography}

\end{document}